\documentclass{aa}
\usepackage{graphicx}
\usepackage{times}
\usepackage{mathptm}
\fontfamily{ptmcm}\selectfont
\newcommand{\mincir}{\raise -2.truept\hbox{\rlap{\hbox{$\sim$}}\raise5.truept
\hbox{$<$}\ }}
\newcommand{\magcir}{\raise -2.truept\hbox{\rlap{\hbox{$\sim$}}\raise5.truept
\hbox{$>$}\ }}
\newcommand{\minmag}{\raise-2.truept\hbox{\rlap{\hbox{$<$}}\raise
6.truept\hbox
{$>$}\ }}

\begin{document}

%\thesaurus{04.19.1; 13.18.2; 03.13.2}
	   
\title{The ATESP Radio Survey}

\subtitle{IV. Optical Identifications and Spectroscopy in the EIS-A 
Region\thanks{Based on observations collected at the European Southern 
Observatory, Chile, under the ESO program identifications 62.O-0883, 
63.O-0467(A) and 64.O-0258(A).}}

\author{I. Prandoni \inst{1,2}
	\and L. Gregorini \inst{1,3}
	\and P. Parma \inst{1}
	\and H.R. de Ruiter \inst{1,4}
	\and G. Vettolani \inst{1}
        \and A. Zanichelli \inst{1,5}
	\and M.H. Wieringa \inst{6}
	\and R.D. Ekers \inst{6}
        }
\offprints{I. Prandoni \\
\email{prandoni@ira.bo.cnr.it}}

\institute{Istituto di Radioastronomia, CNR, Via\,Gobetti 101, I--40129, 
Bologna, Italy
\and Dipartimento di Astronomia, Universit\`a di Bologna, via Ranzani 1,
I--40127, Bologna, Italy
\and Dipartimento di Fisica, Universit\`a di Bologna, Via Irnerio 46,
I--40126, Bologna, Italy
\and Osservatorio Astronomico di Bologna, Via Ranzani 1, I--40127, 
Bologna, Italy
\and Istituto di Fisica Cosmica ``G. Occhialini'', via Bassini 15, 20133 Milano, Italy
\and Australia Telescope National Facility, CSIRO, P.O. Box 76, Epping, 
NSW2121, Australia
}

\date{Received 22 December 2000 / Accepted 1 February 2001}

\titlerunning{THE ATESP Radio Survey. IV}
\authorrunning{I. Prandoni et al.}

\abstract{
This paper is the fourth of a series reporting the results 
of the ATESP radio survey, which was made at 1.4~GHz with the Australia 
Telescope 
Compact Array. The survey consists of 16 radio mosaics with 
$\sim 8\arcsec \times 14\arcsec$ resolution and uniform sensitivity 
($1 \sigma$ noise level $\sim 79$ $\mu$Jy) over the region covered by the ESO 
Slice Project redshift survey ($\sim 26$ sq.~degrees at 
$\delta \sim -40\degr$). 
The ATESP survey has produced a catalogue of 2967 
radio sources down to a flux limit of $\sim$ 0.5 mJy ($6 \sigma$). \\
In this paper we present the optical identifications over a 3 sq.~degr. 
region coinciding with the Patch A of the public ESO Imaging Survey 
(EIS). In this region deep photometry and 95\% complete object catalogues 
in the I band are available down to $I\sim 22.5$.
These data allowed us to identify 219 of the 386 ATESP sources present in the
region. This corresponds to an identification rate of 
$\sim 57\%$. For a magnitude limited sample of 70 optically identified 
sources with 
$I<19.0$ we have obtained complete and good quality spectroscopic data at 
the ESO 3.6 m telescope at La Silla. This data allowed us to get redshift 
measurements and reliable spectroscopic classification for all sources (except
one). \\
From the analysis of the spectroscopic sample we find that the 
composition of the faint radio source population abruptly changes going 
from mJy to
sub--mJy fluxes: the early type galaxies largely dominate the mJy population
($60\%$), while star forming processes become important in the sub--mJy 
regime. Starburst and post-starburst galaxies go from 13\% at $S\geq 1$ mJy
to 39\% at $S<1$ mJy. Nevertheless, at sub--mJy fluxes, early type galaxies 
still constitute a significant fraction (25\%) of the whole population. 
Furthermore we show that, due to the distribution of radio-to-optical ratios, 
sub--mJy samples with 
fainter spectroscopic follow-ups should be increasingly sensitive to the 
population of early type galaxies, while a larger fraction of star-forming 
galaxies would be expected in $\mu$Jy samples.
We compare our results with others obtained from studies of sub--mJy
samples and we show how the existing discrepancies can be explained in terms 
of selection effects. 
}

\maketitle

\section{Introduction}\label{sec-intr}

Deep radio surveys have clearly indicated that normalized radio counts 
show a flattening below a few mJy; the first clear indications of such a 
flattening date back to the pioneering 
deep ($S_{1.4 \; \rm{GHz}} < 1$ mJy) radio surveys undertaken in the early 
eighties (see  
Condon 1984, Kellerman et al. 1987; Windhorst et al. 1990 for reviews on 
radio counts available at the time), and have 
been confirmed by deeper and/or larger surveys (mostly at 1.4~GHz) carried 
out in the last decade. The counts derived from the ATESP survey, which
covers an area 8 times larger than the previous surveys 
(Prandoni et al. 2001, paper III in this series), indicate that 
the upturn actually occurs at $S\la 1$ mJy. 

This change of slope has been interpreted as being due to the presence of 
a new population of radio sources (the so-called sub--mJy population) which
does not show up at higher flux densities.
Classical radio sources, powered by active galactic nuclei (AGN) and 
typically hosted by giant ellipticals and quasars are known to dominate at
high flux densities ($99\%$ above 60 mJy, Windhorst et al. 1990), but their
contribution steadily decreases going to fainter fluxes. At 
$0.1< S_{1.4 \; \rm{GHz}} < 1$ mJy, the median angular size of radio sources
shrinks to $<3\arcsec$ and the sources are predominantly identified with
blue galaxies (see e.g. Kron et al. 1985; Windhorst et al. 1985; Windhorst
et al. 1990). These systems often appear disturbed, in poor groups, and/or
part of merging galaxy systems. This evidence, coupled with the well-known
tight far infrared/radio correlation found for many of these objects at lower
redshift, as well as optical spectra with HII-like signatures similar to 
those of star-forming IRAS galaxies (Benn et al. 1993), suggested that the
radio emission in these galaxies is the result of active star-formation 
($1-100$ M$\odot$/yr). 

Unfortunately, the optical identification work and subsequent spectroscopy,
needed to investigate the nature of the radio sources, are both
very demanding in terms of telescope time, since faint radio sources have 
usually very faint optical counterparts. 
Typically, no more than $\sim 50-60\%$ of  the radio sources in sub--mJy 
samples have been identified on optical images. Only in the $\mu$Jy survey
of the Hubble Deep Field, 80$\%$ of the 111 radio sources 
detected have been identified (Richards et al. 1999).  
On the other hand, the typical fraction of spectra available is only 
$\sim 20\%$. Due to the long integration times needed for detection, deeper 
spectroscopy programmes have been undertaken only for very
small sub--mJy radio samples and never went to completion.  
The best studied sample is the Marano 
Field ($\sim 0.3$~sq.~degr.), where $50\%$ of the sources have spectra 
(Gruppioni et al. 1999), which allowed to determine spectral type and 
redshift for 29 galaxies. 

When discussing results about the nature and the evolution of faint radio
sources, the numbers reported above must be kept in mind. Conclusions are,
in fact, limited by the low identification rate and biased by the fact 
that only the brightest optical counterparts have spectral information. 
This explains why, despite the large observational efforts, 
the true nature of the sub--mJy population is not yet well established. 
Today we know that the sub--mJy population 
is composed by different classes of objects 
(AGN, star-forming galaxies, normal elliptical and spiral galaxies), 
but the relative fractions are still uncertain. 
Gruppioni et al. (1999) found that $\sim 50\%$
of the observed sources show the typical absorption spectra of 
early type galaxies, while late-type galaxies account for 32\% of the 
whole population. This result is in contrast with previous ones, where 
a predominance of star-forming galaxies was found (e.g. in Benn et al., 
$47\%$ of the sources are associated to late-type galaxies). 

Recently the Phoenix survey (Hopkins et al. 1998) has given an important 
contribution to the knowledge of the faint radio population. 
Georgakakis et al. (1999) obtained spectroscopic data for 246 objects with 
$S_{1.4}>0.15$ mJy and $m_R<22.5$, corresponding to about 40\% of the 
optically identified sample. The large numbers involved
allowed the separation, on a reliable statistical basis, the mJy and sub--mJy 
regimes. 
According to their spectroscopic classification, they found that
most of the emission-line sources are star-forming galaxies 
(contributing for $\sim 41\%$ of the sub--mJy population) 
and that the absorption-line systems, likely to be ellipticals, 
dominate at flux densities $>1$ mJy (46\%). However absorption-line systems 
are also found at sub--mJy levels ($\sim 22\%$). 
Unfortunately a significant 
fraction of objects (almost $30\%$) could not be classified due to poor quality
of the spectra, making the quoted numbers still quite uncertain.
 
It is therefore clear that the availability of large faint radio samples
with {\em complete} (good quality) spectroscopy is critical
in order to fully assess the nature and evolution of the mJy and sub--mJy 
radio sources.
A significant advantage for this kind of study is obviously provided
by a faint radio survey with deep photometry already available (possibly 
multicolor). The region we have selected fulfills these requirements at least 
partially. It consists
of the overlap between the ATESP survey (Prandoni et al. 2000a, Paper I) and 
the EIS patch A photometric survey (Nonino et al. 1999).   

This paper is organized as follows: in Sect.~\ref{sec-optid} 
we describe the ATESP-EIS 
sample; in Sect.~\ref{sec-spectra} we define the complete 
sample for which we have obtained spectroscopy and we report the 
observations, the reduction of the spectra and the spectral
classification criteria used. The ATESP-EIS sample properties are described
in Sect.~\ref{sec-ourpopul}. 
The mJy and sub--mJy population properties as
deduced from our analysis are discussed in Sect.~\ref{sec-popul}. 
Conclusions are given in Sect.~\ref{sec-concl}.

Through this paper we will use $H_0 = 100$ km s$^{-1}$ Mpc$^{-1}$ and
q$_0 = 0.5$.

\section{The ATESP--EIS Sample: Optical Identifications}\label{sec-optid}

Vettolani et al. (1997) made a deep redshift survey (referred to as ESP) in 
two strips of $22^{\circ}\times 1^{\circ}$  and $5^{\circ} 
\times 1^{\circ}$ at $\delta \sim -40\degr$. 
They obtained photometry and spectroscopy for 3342 
galaxies down to $b_J \sim$ 19.4 (Vettolani et al. 1998). 
The same region was observed at 1.4 GHz with the Australia Telescope Compact
Array. This radio survey (referred to as the ATESP radio survey) 
consists of 16 radio mosaics with $\sim 8\arcsec \times 14\arcsec$ resolution 
and uniform sensitivity ($1 \sigma$ noise level $\sim$79~$\mu$Jy) over the 
region covered by the ESP redshift survey ($\sim 26$ sq.~degrees, see details
in paper I). The ATESP survey has produced a catalogue of
2967 radio sources down to a flux limit ($6 \sigma$) of $\sim$ 0.5 mJy
(see Prandoni et al. 2000b - paper II - and following updates at 
http://www.ira.bo.cnr.it/atesp). 

\begin{figure}[t]
\vspace{-2cm}
\resizebox{\hsize}{!}{\includegraphics{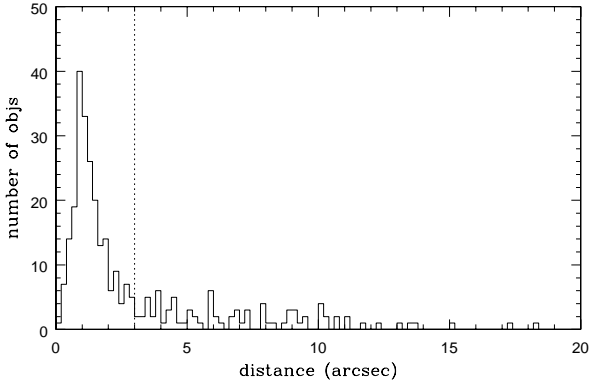}}  
\caption[]{Distribution of the radio-optical positional offsets between each
ATESP radio source in EIS patch A and the nearest EIS object. Only 
radio-optical associations with 
$I< 22.5$ are drawn. 
The vertical dotted line indicates the distance cut-off 
($3\arcsec$) used to define the real radio-optical identifications.
\label{fig-radiooptdist}}
\end{figure}

In a 3.2 sq.~degr. subregion of the $5^{\circ} \times 1^{\circ}$ strip lies 
the EIS (ESO Imaging Survey) Patch A, consisting 
of deep images in the I band out of which an object catalogue has 
been extracted.  
Further V band images are available over $\sim$ 1.5 sq.~degr.. 
In order to identify the radio sources present in this area we used the 
EIS Patch~A I-band {\it filtered} catalogue. 
The filtering is required to eliminate truncated objects (f.i. objects 
at the border of the images) and objects with a
significant number of pixels affected by cosmic rays and/or other artifacts
(see the discussion in Nonino et al. 1999, Sect.~5.3). 
This catalogue covers 3 square degrees. 
At I $\sim$ 22.5 the completeness of the EIS catalog is $\sim 95\%$, while it
rapidly drops at fainter magnitudes.
We then decided to consider in our analysis only the radio-optical 
associations found down to $I=22.5$. 
A search circle of $3\arcsec$ radius, centered at each radio
position, was chosen; this turned out to be a good compromise when inspecting
Fig.~\ref{fig-radiooptdist},
where the distance distribution of the radio-optical associations with
$I<22.5$ is presented.
For double and triple radio sources the distance to the nearest optical 
counterpart is computed from the radio centroid, while for
complex radio sources (e.g. sources which cannot be described by a single or 
multiple Gaussian fit) the distance is computed from the radio peak 
position. Since this position does not generally coincide with the position
of the source nucleus (where the optical identification is expected to be) 
we allowed for distances larger than $3\arcsec$ and checked for the actual 
existence of optical counterpart by visual inspection of the radio-optical
finding charts. 
As to the radio morphology, among the 386 sources three are triples
(none identified), three are complex (one identified), twenty-one are doubles
(six identified) and 359 are point-like (212 identified). In summary, 219 of 
the 386 radio sources present in this region have been identified down to 
I=22.5  

The distribution of radio-optical pairs at distances 
$>3\arcsec$ has been used to provide an estimate of the number of 
spurious identifications expected in the identification sample. 
The estimated contamination rate is $\sim 2\%$ (5/219). 

\begin{figure}[t]
\vspace{-2cm}
\resizebox{\hsize}{!}{\includegraphics{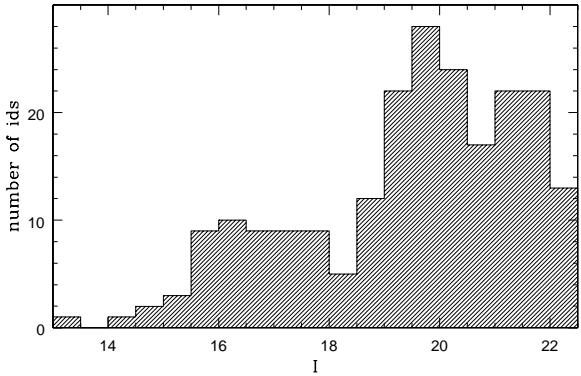}}  
\caption[]{I magnitude distribution for the 219 optically identified ATESP
sources down to $I= 22.5$.
\label{fig-imaghist}}
\end{figure}

Fig.~\ref{fig-imaghist} shows the distribution of the identified radio sources 
as a function of the I-band magnitude. 
The most interesting feature in this distribution
is the strong decrease at magnitudes $I>20$. We point out that this is not
due to incompleteness effects in the optical catalogue but is a consequence
of the physical radio-optical properties of our sources (see discussion in
Sect.~\ref{sec-popul}).

\section{The ATESP-EIS Sample: Spectroscopy}\label{sec-spectra}

From the list of 219 identified radio sources we have extracted a sample of 70
galaxies complete down to I = 19.0 and we have obtained spectra for all 
objects except the double radio source ATESP J224750-400143.
The spectra were taken in three different runs (October 1998, 
September and October 1999) with EFOSC2, mounted on the ESO 3.6 meter 
telescope at La Silla, Chile. In the first run (October 1998) we used 
Grism number 11, which 
gives a dispersion of 2.12 {\AA}/pixel. The scale of EFOSC2 is 0.16 arcsec 
per pixel, corresponding to 13.2 {\AA}/$1\arcsec$ slit. 
The wavelength range covered is 3380--7540 \AA.
In the following runs (September and October 1999) we used 
Grism number 6, which gives 
a similar dispersion (2.06 {\AA}/pixel, corresponding to 
12.9 {\AA}$/1\arcsec$). This grism was chosen because it is more efficient 
in the red part of the wavelength range covered (3860 -- 8070 \AA) and allows
to detect the H$\alpha$ line up to larger redshifts.
Typical exposure times range from 15 minutes for the brightest galaxies
($I<16.5$) to 1 hour for the faintest galaxies ($I>18.5$).
For 14 objects, spectra were already available. They had 
been taken with the
multi-fiber spectrograph Optopus (also at the ESO 3.6 m telescope) by the 
ESP team (Vettolani et al. 1998). These spectra have a dispersion of 
4.5 {\AA} per pixel over the $\lambda$ range 3750--6150 \AA. 

The reduction of the spectra was done with the IRAF package: we followed
the standard procedures of bias subtraction, flat field correction, object 
extraction from the two-dimensional EFOSC images, sky subtraction and 
subsequent
wavelength and flux calibration. In slightly over one half of the spectra 
emission lines were present (typically at least [OII] or H$\alpha$, but often 
other lines too); the elliptical galaxies are mostly without emission lines 
and the redshift is based on absorption lines like the Ca H and K lines, the 
G band, the Mg triplet at $\sim 5180$ \AA, and the Na D doublet. The obtained 
redshifts are helio-centric. The rms 
uncertainty in the redshifts (always determined from at least three lines) is 
of the order of 100 km/s, in accordance with the resolution of the
spectra. 
The high signal-to-noise ratio of these spectra permits an unambiguous
spectral classification of the whole sample. 
Classification of the spectral types is given in the next section.

\begin{figure*}[t]
\resizebox{\hsize}{!}{\includegraphics{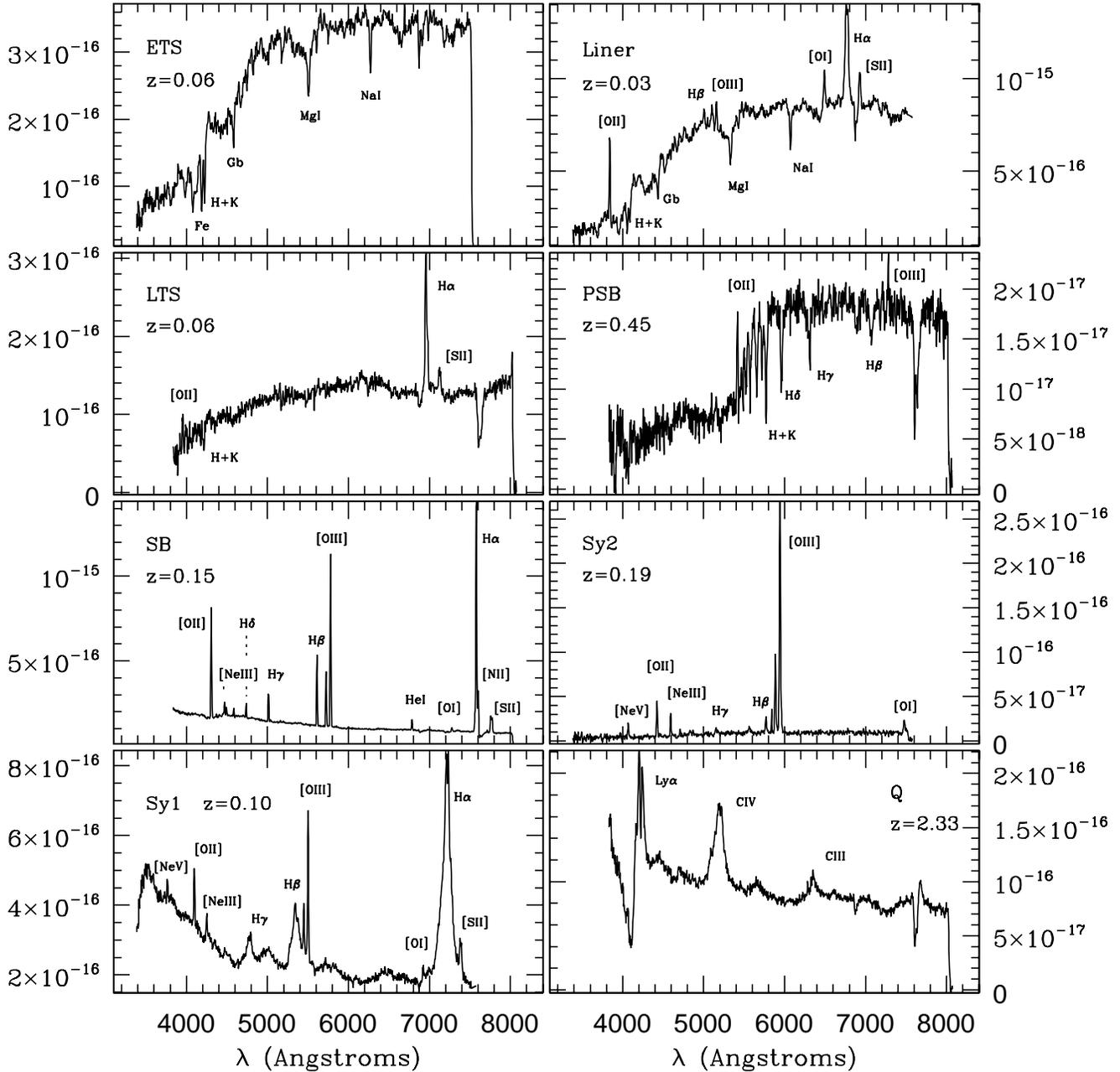}}  
\caption[]{Examples of the different spectral classes defined in the text.
All spectra are redshifted.
\label{fig-templates}}
\end{figure*}

% TABLE 1
\begin{table*}[t]
\caption[]{Optical and radio properties of the complete spectroscopic sample 
\label{tab-list}}
\scriptsize
\begin{tabular}{lrrrrllccl}
\hline\hline\noalign{\smallskip}
\multicolumn{1}{c}{Name} & \multicolumn{1}{c}{$S_{1.4}$} &  
\multicolumn{1}{c}{I} & \multicolumn{1}{c}{V} & 
\multicolumn{1}{c}{$b_j$} & \multicolumn{1}{c}{$z$} & 
\multicolumn{1}{l}{Emission Lines} &
\multicolumn{1}{c}{Spectral} &\multicolumn{1}{c}{Gr.}  
& \multicolumn{1}{l}{Notes}\\
\multicolumn{1}{c}{} & \multicolumn{1}{c}{(mJy)} & \multicolumn{4}{c}{} &
\multicolumn{1}{c}{} & \multicolumn{1}{c}{Class} & \multicolumn{1}{c}{\#} &
\multicolumn{1}{c}{}\\
\hline\noalign{\smallskip}
ATESP J223503-394900&	7.42&	 17.67&       &  19.80 &  0.1957 & [OII], [OIII]&  LTS & 11 & \\
ATESP J223542-402627&	0.71&	 16.63& 18.17 &  18.81 &  0.1665 &  &  LTS & 6 & \\
ATESP J223559-393532&	0.74&	 15.29&       &  17.26 &  0.1071 & &  ETS & & 
ESP spectrum\\
ATESP J223601-393705&	0.94&	 17.87&       &  19.26 &  0.1690 & [OII], [OIII], [OI], H$\alpha$, [SII] &  PSB &  6 &  Signs of SF activity\\
ATESP J223602-401923&	2.16&	 15.91&       &  18.03 &  0.1675 & [OII], H$\beta$, [OIII]&  PSB & 11 & \\
ATESP J223603-400420&	2.96&	 18.57&       &        &  0.5297 & &  ETS & 6 & \\
ATESP J223606-402609&	1.86&	 16.46& 18.06 &  19.26 &  0.2053 & &  ETS & 11 &\\
ATESP J223620-401714&   17.98&	 18.25& 20.77 &        &  0.5075 & &  ETS & 6 & \\
ATESP J223634-402740&	1.13&	 16.27& 17.73 &  18.40 &  0.1409 & [OII] & PSB  & & ESP spectrum\\
ATESP J223636-400444&	0.69&	 16.89& 18.63 &  19.74 &  0.2782 & &  ETS & 6 &\\
ATESP J223656-402554&	3.96&	 17.58& 19.48 &  19.74 &  0.2785 & [OII] &  LTS & 11 &\\
ATESP J223704-393144&   21.96&	 14.05&       &        &  0.0641 & &  ETS & 11 &\\
ATESP J223733-402141&	0.70&	 16.96& 18.25 &  18.99 &  0.1519 & [OII], H$\beta$, [OIII] &  Sy2 & & ESP spectrum \\
ATESP J223747-393612&	0.76&	 17.63&       &  20.14 &  0.2508 & [OII]&  LTS &  6 &\\
ATESP J223812-394018&   15.12&	 16.58&       &  19.10 &  0.2506 & &  ETS & 11 & \\
ATESP J223845-395343&   13.56&	 18.12&       &        &  0.4541 & &  ETS & 6 &\\
ATESP J223848-393345&	0.77&	 15.82&       &  17.29 &  0.0651 & [OII], [OIII]&  LTS & & ESP spectrum\\
ATESP J223916-394125&	2.05&	 16.94&       &  18.95 &  0.1771 & &  ETS & 11& \\
ATESP J224011-393753&	1.58&	 15.08&       &  16.56 &  0.0645 & [OII],
H$\beta$, [OIII]&  LTS  & & ESP spectrum\\
ATESP J224019-394558&	3.20&	 18.00&       &  21.53 &  0.3393 & [OII], 
[OIII]&  LTS & 6 & \\
ATESP J224106-400400&	2.82&	 18.88& 21.05 &        &  0.4662 & &  ETS  &6 &\\
ATESP J224115-394817&	1.79&	 14.72&       &  16.63 &  0.0633 & [OIII] &  LTS & & ESP spectrum \\
ATESP J224120-394844&	7.95&	 18.91&       &        &  0.5948 & &  ETS & 6& \\
ATESP J224139-394150&	1.11&	 17.22&       &  19.14 &  0.1961 & [OII], 
[NeIII], H$\delta$, H$\gamma$, H$\beta$, [OIII], [OI], H$\alpha$& Sy1 &6 &\\
ATESP J224139-394857&	2.58&	 18.34&       &        &  0.1871 & [NeV], 
[OII], [NeIII], H$\gamma$, H$\beta$, [OIII], [OI] & Sy2 & 11 &\\
ATESP J224147-394855&	1.93&	 16.89&       &  18.73 &  0.1851 & [OII] &  LTS & 11 &\\
ATESP J224151-400634&	3.33&	 18.69& 21.27 &        &  0.5032 & &  ETS & 6 &\\
ATESP J224157-402505&	9.70&	 18.13& 20.35 &  20.82 &  0.4265 & &  ETS & 11 &\\
ATESP J224227-395623&   28.12&	 18.54&       &        &  0.5876 & &  ETS  & 6&\\
ATESP J224240-400625&	8.46&	 17.07& 18.54 &  19.55 &  0.2144 & &  ETS & 11 &\\
ATESP J224247-395539&	0.60&	 16.15&       &  17.84 &  0.0604 & [OII], 
H$\beta$, H$\alpha$, [SII] &  LTS & 6 &\\
ATESP J224311-393751&	1.85&	 15.80&       &  17.88 &  0.1263 & &  PSB & 11 &\\
ATESP J224314-400255&   43.17&	 13.42&       &  15.23 &  0.0304 & [OII], H$\beta$, [OIII], [OI], H$\alpha$, [SII]&  L  & 11&\\
ATESP J224322-400839&	0.72&	 17.61& 18.63 &  19.33 &  0.2150 & [OII], H$\gamma$, H$\beta$, [OIII], H$\alpha$, [SII]&  SB & 6 &\\
ATESP J224325-395519&	0.71&	 18.23&       &  20.11 &  0.3267 & [NeV], 
[OII], [OIII] &  LTS & 6 &\\
ATESP J224327-395904&	1.15&	 18.65&       &  21.34 &  0.2846 & &  ETS  & 6&\\
ATESP J224342-400759&	0.61&	 16.06& 17.54 &  18.30 &  0.2164 & &  PSB & &ESP spectrum \\
ATESP J224417-394307&	5.61&	 17.30&       &  19.89 &  0.2832 & &  ETS  & 11&\\
ATESP J224422-393846&	0.57&	 17.13&       &  18.88 &  0.0975 & [OII],
H$\beta$, [OIII]&  SB & &ESP spectrum \\
ATESP J224426-401916&	1.93&	 17.40& 18.84 &  19.58 &  0.2144 & &  ETS & 11 &\\
ATESP J224448-394635&	0.54&	 17.40&       &        &  0.1549 & [OII], 
[NeIII], H$\delta$, H$\gamma$, H$\beta$, [OIII], HeI, [OI], H$\alpha$, [SII]&  SB & 6 &\\
ATESP J224507-394702&	1.24&	 16.60&       &  18.46 &  0.1541 & [OII], 
[OIII], [OI]&  PSB &  11 & Signs of SF activity \\
ATESP J224512-395254&	4.41&	 15.93&       &  18.18 &  0.1387 & &  ETS & &ESP spectrum \\
ATESP J224513-400051&	0.59&	 16.98&       &  19.56 &  0.2468 & &  ETS  &6 & \\
ATESP J224523-393440&	1.19&	 15.21&       &  17.38 &  0.0992 & [NeV], 
[OII], [NeIII], H$\gamma$, H$\beta$, [OIII], [OI], H$\alpha$, [SII]  & Sy1 & 11 &\\
ATESP J224547-400324&   32.83&	 16.22&       &  18.82 &  0.1937 & &  ETS  & 11 &Strong blue  \\
&&&&&&&&& continuum \\
ATESP J224557-393557&	2.11&	 16.20&       &  16.90 &  0.0665 & [OII], 
H$\alpha$, [SII]&  LTS & 11& \\
ATESP J224612-395312&	0.90&	 18.57&       &        &  0.4374 & [OII]&  ETS & 6 &\\
ATESP J224628-401207&	1.35&	 15.92& 17.27 &  18.12 &  0.1259 & &  ETS & 11 &\\
ATESP J224639-393324&	0.91&	 16.66&       &        &  0.1267 & [OII], 
H$\gamma$, H$\beta$, [OIII], [OI], H$\alpha$, [SII]&  SB  & 6 &\\
ATESP J224640-401710&	0.65&	 16.26& 17.68 &  18.49 &  0.1488 & &  ETS & & ESP spectrum \\
ATESP J224654-400108&    5.59&	 17.83&       &        &  0.4011 & &  ETS  & 6 &\\
ATESP J224720-394115&   19.62&	 17.91&       &  20.50 &  0.3274 & [OII]&  LTS  & 11 &\\
ATESP J224729-402001&	0.84&	 16.26& 17.55 &  18.21 &  0.1034 & [OII], H$\beta$, [OIII], H$\alpha$, [SII]&  PSB & 6 & Signs of SF activity \\
ATESP J224750-400143&   13.31&	 18.68&       &  23.15 &         & &  &   & Not observed \\
ATESP J224758-394046&	3.78&	 16.25&       &  18.16 &  0.1313 & &  ETS & &ESP spectrum \\
ATESP J224803-400513&	0.67&	 17.41& 18.00 &  18.04 &  2.3300 & Ly$\alpha$, CIV, CIII &  Q  & 6 &\\
ATESP J224811-395642&	1.15&	 18.73&       &  21.91 &  0.4539 & [OII], [OIII]&  PSB  & 6& \\
ATESP J224821-395707&	1.91&	 15.71&       &        &  0.1332 & [OII], 
[NeIII], H$\delta$, H$\gamma$, H$\beta$, [OIII], [OI], H$\alpha$, [SII]&  SB  &6 &\\
ATESP J224829-394805&	0.57&	 18.95&       &        &  0.4001 & [OII], 
H$\delta$, H$\gamma$, H$\beta$, [OIII]&  SB  & 6 &\\
ATESP J224906-394246&	7.41&	 15.84&       &  17.96 &  0.1354 & &  ETS  & 11&\\
ATESP J224911-400859&	0.88&	 15.95& 17.03 &  17.62 &  0.0646 &  H$\alpha$, [SII]&  LTS & 6 &\\
ATESP J224913-394651&	0.58&	 18.82&       &        &  0.5457 & [OII]&  ETS &  6 &\\
ATESP J224941-395146&	2.26&	 17.70&       &  20.48 &  0.3282 & &  ETS & 6 & \\
ATESP J224958-395855&	1.52&    17.20&       &  19.51 &  0.2489 & &  ETS & 11& \\
ATESP J225008-400425&	2.88&    16.10& 17.36 &  17.78 &  0.1262 & [OII]&  ETS & &ESP spectrum \\
ATESP J225010-395033&	0.93&    17.10&       &  18.78 &  0.1826 & [OII],
H$\gamma$, H$\beta$, [OIII]&  SB  & &ESP spectrum\\
ATESP J225027-394448&	1.38&    17.50&       &  19.85 &  0.3018 & &  ETS &11 &  \\
ATESP J225029-400332&	0.75&    18.81&       &  21.48 &  0.5395 & [OII]&  LTS & 6 & \\
ATESP J225033-394646&	3.16&    14.72&       &  15.23 &  0.0558 & [OII],
H$\beta$, [OIII]&  Sy2 & &ESP spectrum \\
\hline\hline
\end{tabular}
\end{table*}
\normalsize

\subsection{Spectral Classification}\label{sec-discuss}

The spectral classification criteria we used are based on visual inspection
of the spectra and were followed, when necessary, by measurements of line 
fluxes (and/or
equivalent widths) to be used in diagnostic diagrams as those presented
by Baldwin et al. (1981) or Rola et al. (1997). However, the use of diagnostic 
diagrams is rather limited here due to the low resolution of the spectra 
(e.g. $H\alpha$ and [NII] lines cannot be separated) and to the poorer 
signal-to-noise ratio in faint object spectra.
We used the classification scheme originally proposed by Morgan and Mayall 
(1957) and revised by Kennicutt (1992), which is basically followed also 
by Georgakakis et al. (1999)  and Gruppioni et al. (1999) in the study of 
faint 
radio galaxies. However, we have classified the post-starburst 
galaxies separately. The spectral classes we use are:

{\it Early Type Spectrum (ETS)}
-- Galaxies whose spectra show absorption lines coming from a mixture 
of late type stars 
characterizing ellipticals and S0s or bulge dominated spirals. 
The subtle difference between ellipticals and early spirals (especially an 
increase of the blue continuum), is not appreciable in our 
low resolution spectra.
Faint [OII] or $H\alpha + [NII]$ emission lines may be present 
with equivalent width less than 3 Angstroms.

{\it Starburst (SB)} -- We classify as starburst those 
galaxies that show spectra typical 
of moderate to high excitation HII regions (see McCall et al. 1985 for 
examples of extra-galactic HII regions) or similar to the spectrum of NGC3310 
(Kennicut 1992).

{\it Post--starburst (PSB)}
-- We find two types of post-starburst among our galaxies spectra.
a) K+A galaxies which are characterized by strong $H\delta$ absorption 
superposed on a typical ETS spectrum with no emission lines. 
Higher order Balmer absorptions are often present implying the presence of a 
significant A star population which comes from a burst 0.1-1 Gyr old. 
b) Galaxies with spectra showing the simultaneous presence
of Balmer absorptions and a moderate OII line (E+A galaxies, see Zabludoff 
et al. 1996).
Other emission lines can be present. This is indicative of a 
somewhat younger burst or possibly highly reddened still active burst.

{\it Late Type Spectrum (LTS)} -- Galaxies with spectra similar to those of 
late spirals (Sb, Sc, Scd, Sd) which are characterized (with respect to 
ellipticals) by a) bluer stellar continuum, b) less prominent 
absorption lines. Emission lines are often present  
(typically [OII], [OIII], and/or $H\alpha + [NII]$) 
but only in a few cases are
prominent. We notice that in a number of cases it is very difficult to 
assess the prominence of the bulge with respect to the disk. Such objects
have been included in the LTS class. As a consequence, 
this class is probably contaminated by earlier-type spectra, e.g. by 
objects with intermediate properties between the LTS and the ETS classes.

{\it  AGN} -- We group here objects with evident characteristics of 
either Seyfert~1 (Sy1) or Seyfert~2 (Sy2) or quasar (Q) spectra.

{\it Liner (L)} -- There is only one evident liner (as defined in Heckman 1980)
characterized by prominent emission lines from low ionization species 
(e.g. [OI]).

Examples of the spectral classes are illustrated in Fig.~\ref{fig-templates}.
In Table~\ref{tab-list} we present the complete spectroscopic sample: 
the ATESP name is listed in
Col.~1; the total flux for extended radio sources or the peak flux for
point-like sources is given in Col.~2. Fluxes (1.4 GHz) are corrected for 
systematic 
underestimations due to smearing and clean bias (see Paper II for details); 
I and V magnitudes from the EIS
catalogue are given in Cols.~3~and~4, respectively; the $b_j$ magnitude given
in Col.~5, was obtained using the ROE/NRL COSMOS
UKST Southern Sky Object Catalog (Yentis et al. 1992) at the web site of the 
US Naval Observatory; redshift is in Col.~6, prominent emission 
lines\footnote{H$\alpha$ is always intended as H$\alpha$+[NII].}, whenever 
present, and the spectral class are given in
Cols.~7 and~8, respectively. The last two columns report the Grism 
information (Col.~9) and some notes about individual 
spectra (Col.~10).

\section{The ATESP-EIS Sample: Properties}\label{sec-ourpopul}

\begin{table}[t]
\small
\caption[]{Composition of the ATESP-EIS sample \label{tab-comp}} 
%\begin{tabular}{|l|rr|rr|rr|}
\begin{tabular}{lrrrrrr}
\multicolumn{7}{l}{}\\
\hline
\multicolumn{1}{l}{Spectral Class}  & \multicolumn{2}{c}{All} &
\multicolumn{2}{c}{$S\geq 1$ mJy} & 
\multicolumn{2}{c}{$S< 1$ mJy} \\
\multicolumn{1}{c}{}  
& \multicolumn{2}{c}{\footnotesize{$N \; \; \; \; \; \;  \; \;  (\%)$}} &
\multicolumn{2}{c}{\footnotesize{$N \; \; \; \; \; \;  \; \;  (\%$)}} & 
\multicolumn{2}{c}{\footnotesize{$N \; \; \; \; \; \;  \; \;  (\%)$}} \\
\hline 
  & & & & & & \\
 Early Type Spectrum  & 33 & (48)  & 27 & (60) & 6 & (25)   \\
+ Liner & & & & & & \\
 AGN    & 6 & (9)  & 4 & (9)  & 2 & (8)  \\
 Late Type Spectrum & 15 & (22)  & 9 & (20) & 6 & (25)\\
 Starburst + post-SB & 15 & (22)  & 6 & (13) & 9 & (39) \\
 All & 69 &  & 46 & (67) & 23 & (33) \\
  & & & & & & \\
\hline\hline 
\end{tabular}
\end{table}

The faint radio source composition resulting from the ATESP-EIS
spectroscopic sample is presented in Table~\ref{tab-comp}, where
sub--mJy and mJy regimes have been considered separately. We notice that
the good quality of the spectra allowed us to classify all objects.
Our data clearly show that the AGN contribution  does not
significantly change going to fainter fluxes ($8-9\%$). The same is
true for late type (LTS) objects (which go from 20\% to 25\%). This can be due
to the fact that this spectral class is probably contaminated by objects
with earlier type spectra (see discussion 
in Sect.~\ref{sec-discuss}). More meaningful is therefore a direct comparison 
between the class of early type (ETS) galaxies and the starburst + 
post-starburst (SB + PSB) one. 
Early type galaxies largely dominate (60\%) the mJy
population, while star-formation processes become important in the sub--mJy
regime: SB and post-SB galaxies go from 13\% at $S\geq 1$ mJy to 39\% at
$S<1$ mJy.
Nevertheless, at sub--mJy fluxes, early type galaxies still constitute a
significant fraction (25\%) of the whole population. 

\begin{figure}[t]
\resizebox{\hsize}{!}{\includegraphics{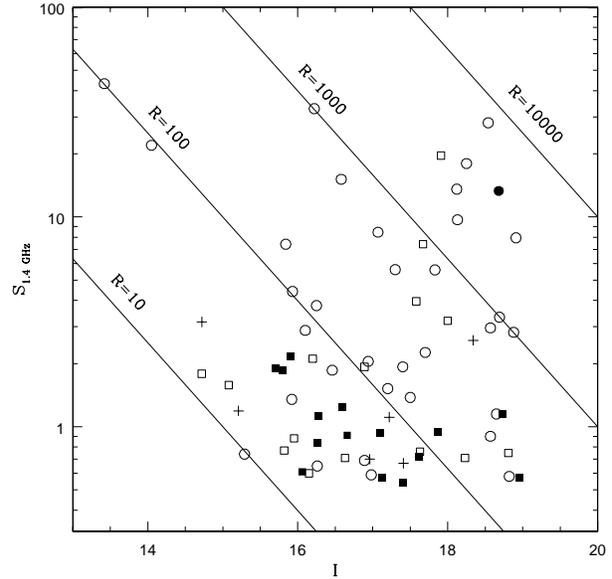}}  
\caption[]{1.4 GHz flux density (in mJy) versus I magnitude; 
lines represent constant values of the radio to optical ratio $R$.
Symbols represent different spectral classes: Early Type Spectrum + Liner
(empty circles), Late Type Spectrum (empty squares), Starburst + post-SB 
(filled squares) and AGN (crosses). 
Also drawn is the single object for which 
spectroscopy is not available (filled circle).
\label{fig-ratio}}
\end{figure}

\begin{figure}[t]
\resizebox{\hsize}{!}{\includegraphics{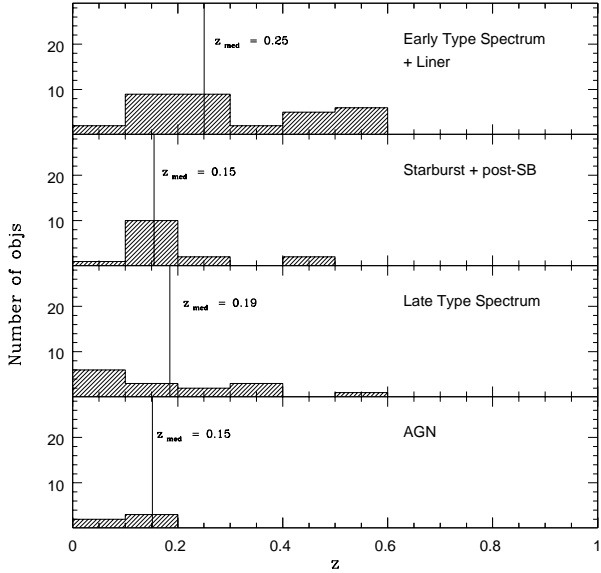}}  
\caption[]{Redshift distribution for the ATESP spectroscopic sample. 
Different spectral classes are drawn separately. From Top
to Bottom: Early Type Spectrum + Liner; Starburst + Post-SB;
Late Type Spectrum; AGN (the quasar at z=2.33 is not drawn and not considered 
in the computation of the median redshift).
\label{fig-zeta}}
\end{figure}
 
In Fig.~\ref{fig-ratio} we plot the radio flux densities against the 
I magnitudes for the whole sample. Superimposed are the lines representing
constant values for the radio-to-optical ratios, defined following Condon 
(1980) as 
\begin{equation}\label{eq-ratio}
R=S \cdot 10^{0.4(I-12.5)} \, ,
\end{equation} 
where $S$ is the 1.4 GHz flux in mJy and $I$ is the magnitude. 
In Fig.~\ref{fig-ratio} the sudden change in the 
radio source composition going from the mJy to the sub--mJy regime is 
immediately evident. From a closer inspection, though, it also appears that 
the fraction of sub--mJy sources with early type spectrum becomes 
more important going to fainter magnitudes. 
In fact at sub-mJy fluxes
the number ratio of star-forming (filled squares) over early type 
(empty circles) galaxies is $2:1$ for $I<18$ (8 SB/PSB galaxies against 
4 ETS galaxies), while a reversed ratio is found for $I>18$ (2 ETS against 
1 SB).  
The latter result, even though based on a very small number of objects,
may indicate that star-forming galaxies actually dominate the sub--mJy 
population only at bright magnitudes, as also found by Gruppioni et al.
(1999) in the radio-optical study of the Marano Field (see discussion in 
Sect.~\ref{sec-popul}). 

\begin{figure}[t]
\resizebox{\hsize}{!}{\includegraphics{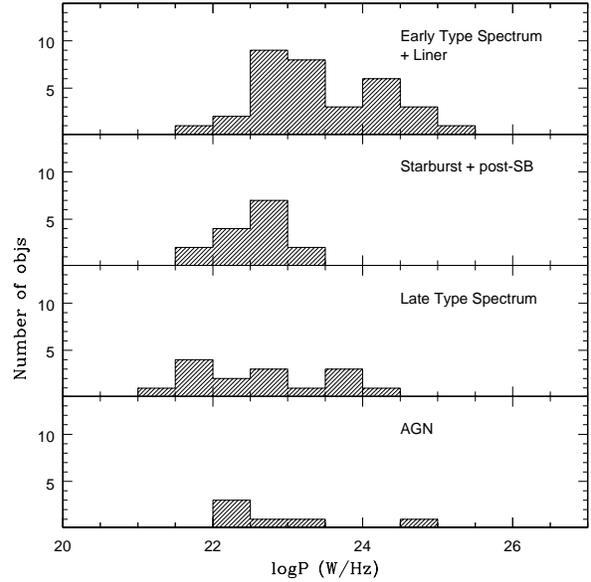}}  
\caption[]{Radio power distribution for the ATESP spectroscopic sample. 
Different spectral classes are drawn separately. From Top
to Bottom: Early Type Spectrum + Liner; Starburst + Post-SB;
Late Type Spectrum; AGN.
\label{fig-logp}}
\end{figure}

The fact that star-forming galaxies dominate our sub--mJy 
sample only at bright magnitudes is
reflected also in the redshift distribution of the different spectral
classes (shown in Fig.~\ref{fig-zeta}). 
The median redshift distribution of the whole sample is z=0.20; but
starburst and post-starburst galaxies are nearer than early type galaxies. 
Such a 
trend was also found in previous studies of sub--mJy (Gruppioni et al. 1999) 
and mJy (Magliocchetti et al. 2000) samples.

In Fig.~\ref{fig-logp} we show the radio power distribution of the different 
spectral classes. These distributions reflect the radio luminosity properties 
of the different classes. In particular, all starburst and post-starburst 
galaxies have powers $P<10^{23.5}$ W/Hz, while the 
classes of early type objects and AGNs 
extend up to $P\simeq 10^{25.5}$ W/Hz. Both results are in very good 
agreement with the luminosity functions derived from local samples 
of radio sources: star-forming galaxies mostly contribute at radio 
powers $P<10^{23}$ W/Hz (and become negligible in number at $P>10^{24}$ W/Hz),
while early type galaxies show a much flatter distribution in radio power,  
extending up to $P> 10^{26}$ W/Hz (e.g. Condon 1989). 
The class of late type objects cover a larger range in radio power than the 
class of starbursts and post-SBs, supporting the hypothesis that this class 
can be contaminated by earlier type objects.

Figs.~\ref{fig-ratio} and~\ref{fig-logp} indicate a possible physical 
interpretation for the different composition found going from mJy to sub--mJy 
fluxes, and its possible dependence on the optical magnitude. 
Star-forming galaxies, characterized by weak intrinsic radio 
emission, have low radio-to-optical 
ratios (typically $10<R<100$, see Fig.~\ref{fig-ratio}), while early 
type galaxies cover a much larger
range in radio power, and hence in R, becoming the dominant population at 
$R>100$. 
If this behaviour holds going to fainter fluxes and magnitudes, 
sub--mJy samples with 
fainter spectroscopic follow-ups should be increasingly sensitive to the 
population of early type galaxies, while, at a given optical limit,
a larger fraction of star-forming galaxies would be expected in 
$\mu$Jy samples, as supported by the study of the
$\mu$Jy sources in the Hubble Deep Field, the majority of which seem
to be associated with star-forming galaxies (Richards et al. 1999).

\begin{figure}[t]
\resizebox{\hsize}{!}{\includegraphics{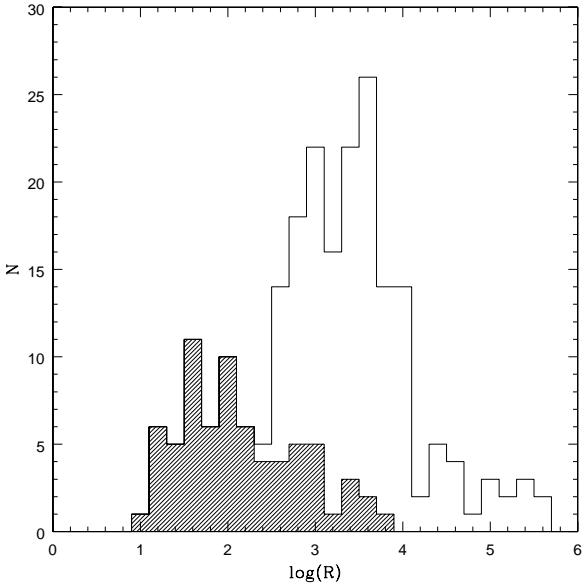}}  
\caption[]{Distribution of the radio-to-optical ratio $R$ 
for the 219 identified galaxies down to $I=22.5$ (empty histogram) and for 
the 70 galaxies with $I<19.0$ of the spectroscopic sample (shaded histogram).
\label{fig-ratiohist}}
\end{figure}

This hypothesis is also supported by Fig.~\ref{fig-ratiohist}, where we
present the distribution in $R$ for both the 
sample of all the radio sources identified down to $I=22.5$ (empty histogram)
and the 70 objects of the spectroscopic sample ($I<19.0$)
(shaded histogram). In fact the 
spectroscopic sample essentially exhausts the objects with  
low radio-to-optical ratio, while the optically fainter objects 
mostly have $R>>100$. This indicates, even accounting for some evolution,  
that they are essentially early type galaxies or AGN's. 
Deeper spectroscopy for the ATESP-EIS sample will be crucial
in order to verify this indication on a reliable statistical basis and to
quantify the importance of evolutionary effects in the faint radio population. 

As a final remark we notice that the position of the only object with no 
spectroscopy available in the flux-magnitude diagram (filled circle at 
$R\sim 4000$), together
with a double-component radio morphology, points toward a nuclear origin
for its radio emission. 

\section{The mJy and sub--mJy Population}\label{sec-popul}

As already discussed in Sect.~\ref{sec-intr}, the results obtained from
the study of faint radio samples are not completely in agreement with each 
other
and therefore the nature of the mJy and sub--mJy radio population is 
not yet conclusively established. This is due to both
the limited size of the faintest radio surveys and to the limited optical 
follow-up available.  

The possibility that selection effects could play an important 
role in this kind of study, was first recognized by Gruppioni et al. (1999)
from the study of the Marano Field. They found a 
higher fraction of early type galaxies than in previous studies: 50\% of the 
sources identified have colours and spectral 
features typical of early type galaxies, against e.g., the 9\% found by Benn 
et al. (1993). According to the authors, this discrepancy 
can very likely be ascribed to the deeper optical magnitude limit 
reached in the Marano Field,
since the fraction of sources identified with early type galaxies abruptly 
increases at magnitudes $B>22$, which is approximately the limit of the 
sample studied by Benn et al. (1993). 

In order to properly compare the results reported in 
the literature, it is therefore very important to take into account the 
differences in limiting flux 
and magnitude of the available radio samples. 

If we assume (as suggested in Sect.~\ref{sec-ourpopul})
that a radio-to-optical ratio $R\simeq 100$ separates the locus where
the star-forming population dominates ($R<100$) from the one dominated by the 
early-type population ($R>100$), we can set a
limit on the brightest limiting magnitude required in the spectroscopy
follow-up, in order to make the radio-optical sub--mJy studies increasingly 
sensitive to the early-type population. 
Inverting Eq.~(\ref{eq-ratio}), we find
\begin{equation}
I_{100} > 17.5 -2.5 \log{S}
\end{equation}
where $I_{100}$ is defined as the magnitude limit corresponding to $R>100$
and depends on the radio flux $S$.   

In Table~\ref{tab-others} (Column~3) 
we report the different values of $I_{100}$ computed assuming
$S=S_{lim}$ ($S_{lim}$ is given in Column~2) 
of the most recent 1.4 GHz mJy and sub--mJy samples 
with radio-optical studies available.
The samples are sorted on increasing values of the quantity $I_{lim}-I_{100}$
($\Delta I$ in Column~5), where $I_{lim}$ is the limiting magnitude of
spectroscopic follow-up in each sample. To make the comparison among the 
different samples easier 
we translated all the limiting magnitudes to I-band values, assuming
indicatively $R-I=0.5$ and $B-I=1.5$. In the following columns we give
the relative fractions for the different classes of objects found in each 
sample. To allow for the many sub-classes defined by different authors 
we grouped the objects in three main classes: Early Type galaxies (E), 
Late Type galaxies (L) and Active Galactic Nuclei (AGN). The Late Type class
contains starburst, post-starburst and normal spirals. The fourth class (O)
contains any other kind of object which cannot be included in the first three
classes. 
Besides a number of objects classified as stars, they are mostly 
unclassified objects due to failed and/or poor quality spectra.  

Under the hypothesis that 
the important issue in determining the composition of the faint radio 
population it is neither the radio flux nor the optical magnitude, but a 
combination of the two quantities, the relative importance of early type
objects should increase with increasing $\Delta I$, that is going through the
table, from the B93 sample (the one studied by Benn. et al. 1993) to the 
Marano Field (MF, Gruppioni et al. 1999). 
Even if the differences among the quoted fractions in the different samples
are not statistically relevant due to the poor number of objects involved, it
is interesting to note that this trend is confirmed 
for both the mJy and the sub--mJy sample. The only exception is represented
by the Phoenix Deep Field (PDF, Georgakakis et al. 1999), where a fraction of 
early type objects smaller than expected is found. It is worth noticing though
that the high percentage of 
unclassified objects makes the PDF result quite uncertain. 
This uncertainty is probably more severe for the class of the early type 
galaxies. 
In fact most of the failed and/or poor quality spectra correspond to objects 
with very faint magnitudes, where many early type objects are 
expected. Another bias against early type objects is  
the fact that pure absorption spectra (typical of early type galaxies) 
are more difficult to obtain than emission line spectra (typical of late type
objects and AGNs). The same uncertainty affects the numbers reported 
for the Benn et al. sample, where the quoted fractions have to be 
considered as lower limits. With this caveat, the expected trend for
a decreasing fraction of Late-type objects at sub--mJy fluxes with increasing
$\Delta I$ is verified.

Very interesting is the fact that the fraction of AGNs  is very similar 
(10-15\%) in all the samples and in both flux regimes. This class is also
probably the least affected by uncertainties due to poor spectroscopy, because
in general AGNs show very strong emission lines. Since these faint radio 
surveys are mostly sensitive to small radio-to-optical ratios ($R<1000$), 
this result can give a reliable constraint on the number density of low 
luminosity AGNs. 

\begin{table*}
\caption[]{Comparison with other 1.4 GHz mJy and sub--mJy samples 
\label{tab-others}}
\begin{tabular}{lclll|ccccc|ccccc}
\hline\hline
\multicolumn{1}{l}{} &  \multicolumn{1}{c}{}
& \multicolumn{3}{c}{}& 
\multicolumn{5}{|c|}{$S\geq 1$ mJy} &\multicolumn{5}{|c}{$S< 1$ mJy}\\
\multicolumn{1}{l}{Survey} & \multicolumn{1}{c}{$S_{lim}$} 
& \multicolumn{1}{c}{$I_{100}$} & \multicolumn{1}{c}{$I_{lim}$} &
\multicolumn{1}{c}{$\Delta I$}  &
\multicolumn{1}{|c}{E} & \multicolumn{1}{c}{L} 
& \multicolumn{1}{c}{AGN} & \multicolumn{1}{c}{O} & 
\multicolumn{1}{c|}{$N_{tot}$} &
\multicolumn{1}{|c}{E} & \multicolumn{1}{c}{L} 
& \multicolumn{1}{c}{AGN} & \multicolumn{1}{c}{O} &
\multicolumn{1}{c}{$N_{tot}$}\\
\multicolumn{1}{c}{} & \multicolumn{1}{c}{(mJy)} & \multicolumn{3}{c}{}& 
\multicolumn{1}{|c}{(\%)} & \multicolumn{1}{c}{(\%)} 
& \multicolumn{1}{c}{(\%)} & \multicolumn{1}{c}{(\%)} & 
\multicolumn{1}{c|}{} &
\multicolumn{1}{|c}{(\%)} & \multicolumn{1}{c}{(\%)} 
& \multicolumn{1}{c}{(\%)} & \multicolumn{1}{c}{(\%)} &
\multicolumn{1}{c}{}\\
\hline
 & & & & & & & & & & & & &  &\\
B93   & 0.1 & $20.0$ & $20.5$ & 0.5  & 33 & 22 & 11 & 33 & 18 & 3 & 54 & 13 
& 30 & 69 \\
FIRST & 1.0 & $17.5$ & $18.1$ & 0.6  & 52 & 24 & 15 & 9 & 46 & - & - & - & - 
& - \\
ATESP-EIS  & 0.5  & $18.3$ & $19.0$ & 0.7 & 60 & 33 & 9 & - & 46 & 25 & 65 & 8 
& - & 23    \\
PDF   & 0.2 & $19.3$ & $21.0$ & 1.7 & 46 & 11 & 11 & 31 & 71 & 22 & 41 & 13 
& 25 & 175  \\
MF  & 0.2 & $19.3$ & $23.0$ & 3.7 & 73 & 9 & 18 & - & 11 & 39 & 43 & 13 & 4 
& 23    \\
 & & & & & & & & & & & & & &  \\
\hline 
\hline
\multicolumn{2}{l}{B93: Benn et al. 1993} &
\multicolumn{5}{l}{FIRST: Magliocchetti et al. 2000} &
\multicolumn{3}{l}{ATESP-EIS: this paper} &
\multicolumn{5}{l}{PDF: Georgakakis et al. 1999}\\
\multicolumn{15}{l}{ MF: Gruppioni et al. 1999 }\\
\end{tabular}
\end{table*}

With this kind of arguments it is also possible to explain the general trend 
for higher fractions of early type objects at mJy fluxes than at sub--mJy 
fluxes: at $S\geq 1$ mJy 
the $R<100$ population is expected at very bright magnitudes 
($I<I_{100}\leq 17.5$), where the existing samples (covering areas from a 
fraction of a sq.~degr. to a few sq.~degr.) are severely volume limited.  

This analysis seems to be able to reconcile, at least 
qualitatively, the discrepancies found in the literature about the nature
of the faint radio sources. In particular it is very promising that
in the two samples where the change of the population with 
the optical magnitude
has been investigated (the ATESP-EIS and the MF) the computed values for 
$I_{100}$ corresponds very well to the magnitude where the change has been
found. In the ATESP-EIS we found an increase of early type objects at $I>18$
(to be compared with $I_{100}=18.3$), while in the MF the same increase 
is found at $R>20$ (see Fig.~4(b) in Gruppioni et al. 1999), roughly 
corresponding to $I>19.5$ (to be compared with $I_{100}=19.3$). 

As a final remark we notice that the strong decrease at faint magnitudes 
in the distribution of identified radio sources as a function of I 
magnitude shown in Fig.~\ref{fig-imaghist} can be explained, when we 
consider that at $I>20$ the ATESP-EIS sample is sensitive only to 
radio sources with $R>500-1000$ at fluxes of the order of $0.5-1$ mJy. 
Such values correspond to early type galaxies and/or AGNs, which are not
the dominant population in optically selected samples like the EIS. 
A similar result was found by Georgakakis et al. (1999) in the PDF sample, 
where the proportion of star-forming galaxies was found to decrease
relative to those of early type systems and AGNs at
magnitudes $R>19.5$. On the other hand, the authors could not draw firm
conclusions from this result due to selection effects and 
incompleteness severely affecting their sample at these magnitudes.  
From our analysis we can argue that the decrement of star-forming galaxies 
found in the PDF sample at $R>19.5$ is probably mostly real, since 
we expect this sample to become increasingly insensitive to star-forming 
galaxies at magnitudes $I>I_{100}=19.3$ (or $R>19.8$).

\section{Conclusions}\label{sec-concl}

We have studied the radio-optical properties of a magnitude limited sample
of 70 radio sources with spectroscopy available ($I<19.0$).
From our analysis we find that the 
composition of the faint radio sources abruptly changes going from mJy to
sub--mJy fluxes: the early type galaxies largely dominate the mJy population
($60\%$), while star forming processes become important in the sub--mJy 
regime. Starburst and post-starburst galaxies go from 13\% at $S\geq 1$ mJy
to 39\% at $S<1$ mJy. Nevertheless, at sub--mJy fluxes, early type galaxies 
still constitute a significant fraction (25\%) of the whole population. 

We argue that the relative fractions found for the two classes of star-forming
and early type galaxies mainly depend on the fact that star-forming
galaxies dominate at low radio-to-optical ratios ($R<100$), while  
early type objects dominate at $R>100$. 
Under this hypothesis it is possible to reconcile the discrepancies 
found from the radio-optical analysis of different sub--mJy radio samples, 
by taking into account the differences in limiting radio flux and optical 
magnitude. 

If this behaviour holds going to $\mu$Jy fluxes and to fainter magnitudes than
reached by current studies, we expect that faint radio samples with 
fainter spectroscopic follow-ups will be increasingly sensitive to the 
population of early type galaxies, with a larger fraction of star-forming 
galaxies in $\mu$Jy samples (given the same optical limiting magnitude). 
In particular, for a given radio sample, it would be possible to 
estimate a critical value for the optical magnitude, corresponding to the 
transition from star-forming to early type galaxy population: at brighter
magnitudes the dominant class would be the one of star-forming galaxies, while
at fainter magnitudes the early type galaxies would become the main class.
This would also correspond to a limit in redshift for the star-forming 
class that populate a given faint radio sample. In other words, studies based 
on the star-forming portion of sub--mJy samples (like f.i. the analysis
of the evolution of the star formation rate with the cosmic time) can give
information up to limited redshifts ($z<0.5$), while larger redshift values 
(up to $z\sim 1$) can be probed only by $\mu$Jy samples. 

In order to verify this hypothesis and check whether evolutionary effects
could play an important role in changing this scenario, complete
optical follow-ups down to very faint magnitudes ($I>25$) for  
statistically relevant faint radio samples are needed. 
To avoid bias and/or
incompleteness effects at the faintest magnitudes it is also very important 
to obtain good quality spectroscopy. 

\begin{acknowledgements}
EIS observations have been carried out using the ESO New Technology Telescope 
(NTT) at the La Silla observatory under Program-ID Nos. 59.A-9005(A) and 
60.A-9005(A). 
I.P. thanks the ESO staff and all the people directly and indirectly involved 
in the EIS Imaging Survey for their hospitality and support, during the period
spent in Garching as a member of the EIS team. \\
The spectroscopy data were taken at the 3.6 m telescope at ESO La Silla 
Observatory under the ESO program identifications 62.O-0883, 63.O-0467(A) and
64.O-0258(A).
We thank the ESP team, and in particular Elena Zucca and Roberto Merighi,
for providing detailed information about the ESP spectra used in this work. 
Roberto Fanti is acknowledged for many useful comments. We are grateful 
to the anonymous referee for a number of useful suggestions. \\
The authors acknowledge the Italian Ministry for University and 
Scientific Research (MURST) for partial financial support (grant 
COFIN 99-02-37). 
\end{acknowledgements}

\end{document}